\documentclass[%
 reprint,
 amsmath,amssymb,
 aps,
pra,
]{revtex4-1}

\usepackage{color}
\usepackage[table]{xcolor}
\usepackage{graphicx}
\usepackage{dcolumn}
\usepackage{bm}

\definecolor{lightlightgrey}{gray}{0.92}

\newenvironment{exer*}
  {\ex}
  {\endex}

\def\be{\begin{equation}}
\def\ee{\end{equation}}
\def\bea{\begin{eqnarray}}
\def\eea{\end{eqnarray}}
\newcommand{\ket}[1]{\mbox{$|#1\rangle$}}
\newcommand{\bra}[1]{\mbox{$\langle#1|$}}
\newcommand{\avg}[1]{\mbox{$\langle#1\rangle$}}

\newcommand{\opd}[2]{\mbox{$\hat{#1}_{#2}^{\dagger}$}}  
\newcommand{\op}[2]{\mbox{$\hat{#1}_{#2}$}}

\newcommand{\aopd}{\hat{a}^\dagger}
\newcommand{\aop}{\hat{a}}

\newcommand{\kappaex}{\kappa_\text{ex}}
\newcommand{\kappain}{\kappa_\text{in}}

\newcommand{\xzp}{x_\text{zp}}

\newcommand{\omegam}{\omega_\text{m}}
\newcommand{\omegao}{\omega_\text{o}}

\newcommand{\Nbnc}{N_\text{rep}}

\usepackage{xspace} 

\newcommand{\hairsp}{\hspace{1pt}} 
\newcommand{\ie}{\textit{i.\hairsp{}e.}\xspace} 

\usepackage[amssymb]{SIunits}
\usepackage{ulem}
\graphicspath{{figures/}}
\usepackage{float}

\begin{document}


\title{Enhancing a slow and weak optomechanical nonlinearity with delayed quantum feedback}

\author{Zhaoyou Wang}%
\email{zhaoyou@stanford.edu}
\author{Amir H. Safavi-Naeini}
\email{safavi@stanford.edu}
\affiliation{%
Department of Applied Physics, and Ginzton Laboratory, Stanford
University, Stanford, California 94305, USA
}%

\date{\today}

\begin{abstract}
One of the central goals of quantum optics is to generate large interactions between single photons. Light interacting with motion in an optomechanical system can sense minute fluctuations in displacement, and also impart a force via radiation pressure. Taken together, these two effects mean that two photons can ``sense'' each other's presence in an interaction mediated by motion. It is accepted that for an optomechanical system to mediate strong interactions between single photons, the mechanical system must respond before the photon is lost ($\omegam > \kappa$), and the radiation pressure force must generate a displacement large enough to change the optical properties of the system ($g_0 > \kappa$). The challenge of achieving this ``vacuum strong coupling'' has prevented experiments from demonstrating single-photon interactions. In this work we show that by adding a coherent feedback channel to a slow mechanical system ($\omegam < \kappa$) that is weakly nonlinear ($g_0 < \kappa$), two spatially separated single photons can be made to effectively interact with each other deterministically through the mechanical motion of a resonator. To numerically analyze our system, we must solve Schr\"odinger's equation for the state of an optomechanical system coupled to a long waveguide feeback channel. We implement a matrix product state approach to keep track of and evolve the complete quantum state of the system in an efficient way. We analyze the process semiclassically and then solve the full quantum dynamics numerically to find a cross-over between the semiclassical and quantum regimes of optomechanics. Finally we analyze the experimental prospects for implementing this protocol.
\end{abstract}

\maketitle

\section*{Introduction}
The interaction between light and motion in cavity-optomechanical systems has enabled sensitive measurements of force and displacement, as well as quantum-optical control over the state of mechanical resonators~\cite{Aspelmeyer2014}. Conversely, an optically-coupled mechanical degree of freedom can lead to interactions between photons mediated by motion. Nonlinear optical effects such as wavelength conversion~\cite{Dong2012,Hill2012,Andrews2014,Lecocq2016}, generation of squeezed light~\cite{Brooks2012,Safavi-Naeini2013,Purdy2014}, and electromagnetically induced transparency~\cite{Weis,Safavi-Naeini2010e} are manifestations of this optical nonlinearity and have been demonstrated in recent experiments. Nonetheless, a central goal of quantum optics and information is to generate nonlinearities that are large at the single-photon level~\cite{Chuang1995}. It has been shown theoretically that it is possible to generate single-photon effective optical nonlinearities with optomechanics, but only with system parameters well outside of our current experimental reach~\cite{Rabl2011,Nunnenkamp2011,Ludwig2012b,Stannigel2012,Lemonde2015,Xuereb2012,Asjad2015}. The essential difficulty is that for a mechanically-induced optical nonlinearity to act on the incident light, the response of the mechanical system must be fast compared to the amount of time the photon spends inside the optical cavity. Moreover, the mechanical system should display a large response to the force induced by a single photon -- so that the cavity properties change significantly due to the presence of the photon -- requiring a highly compliant mechanical system. These two contradicting needs, fast response and large compliance, are at odds with each other and make strong effective photon-photon interactions extremely difficult to achieve in realistic cavity optomechanical systems.

In addition to the difficulty of generating the large Kerr nonlinearities needed, it has been pointed out that in principle, subtle effects due to the multimode nature of a propagating light field make implementation of high-fidelity gates operating on flying photons using large instantaneous Kerr nonlinearities  impossible~\cite{Shapiro2006a}.

Recently, it has been proposed that optomechanical nonlinearities can be enhanced in a fast-cavity system by parametric amplification of mechanical motion~\cite{Lemonde2015} leading to a quantum gates between photons of different frequency. We propose a different way to make strong photonic interactions from weak optical nonlinearities. In this article, we show that by introducing a \textit{coherent delay} to a cavity-optomechanical system with a small optomechanical coupling and slow response, large photon-photon interactions between two temporally separated photon modes propagating in an optical fiber can be achieved. Time-delayed coherent feedback has been discussed for Gaussian states in other contexts~\cite{Pikovski2012,Nemet2016}. Recently a new method based on Matrix Product States (MPS)~\cite{Pichler2016} has been developed that greatly facilitates a full quantum analysis of such interactions. Here, we demonstrate using both a semiclassical argument and full quantum simulations taking into account the  propagating quantum field with its many degrees of freedom, that a {\small{CPHASE}} gate between two temporal modes of the field can be implemented with a cavity-optomechanical system that is in the bad-cavity regime. The coherent delay allows a slow optomechanical system to induce a large effective interaction between the temporally separated photons, greatly relaxing the optomechanical system requirements needed to implement such a gate. The nonlocal nature of the gate also sidesteps a key assumption in the aforementioned impossibility arguments~\cite{Shapiro2006a,Gea-Banacloche2010,Gea-Banacloche2014}.

\section*{Results}
\textbf{System and {\small{CPHASE}} gate protocol.} We consider an optomechanical system coupled to a long waveguide with an end mirror, so that photons can propagate back and forth inside the waveguide, and interact repeatedly with the mechanical resonator by entering the optical cavity. The cavity-optomechanical system is composed of an optical resonator at frequency $\omegao$ with annihilation operator $\hat{a}$, coupled to a mechanical resonator with frequency $\omegam$ and annhililation operator $\hat{b}$. The initial state of the waveguide consists of two temporal modes of the light field with Gaussian profile, each with extent $\tau$. We assume that $1/\tau \ll \kappa$, where $\kappa$ is the optical loss rate, so that each photon can fully enter the cavity and exert a significant impulse onto the mechanical system. During this process, the photon also obtains an uniform phase shift caused by the internal state of the mechanical system. Each impulse is assumed to be nearly instantaneous on time scales relevant to the mechanical oscillator's motion, \ie, $\omegam \ll 1/\tau$. Taken together, these conditions imply that we are operating in the bad cavity regime $\omegam \ll \kappa$.

The slow internal dynamics of the mechanical system as well as the coherent time delay lead to an effective nonlinear photon-photon interaction that is nonlocal in time and also leads to a build up of entanglement between the two temporal modes inside the waveguide. Remarkably, we find that a quantum phase gate can be implemented in this system.

\begin{figure}
  \centering
  \includegraphics[width=0.5\textwidth]{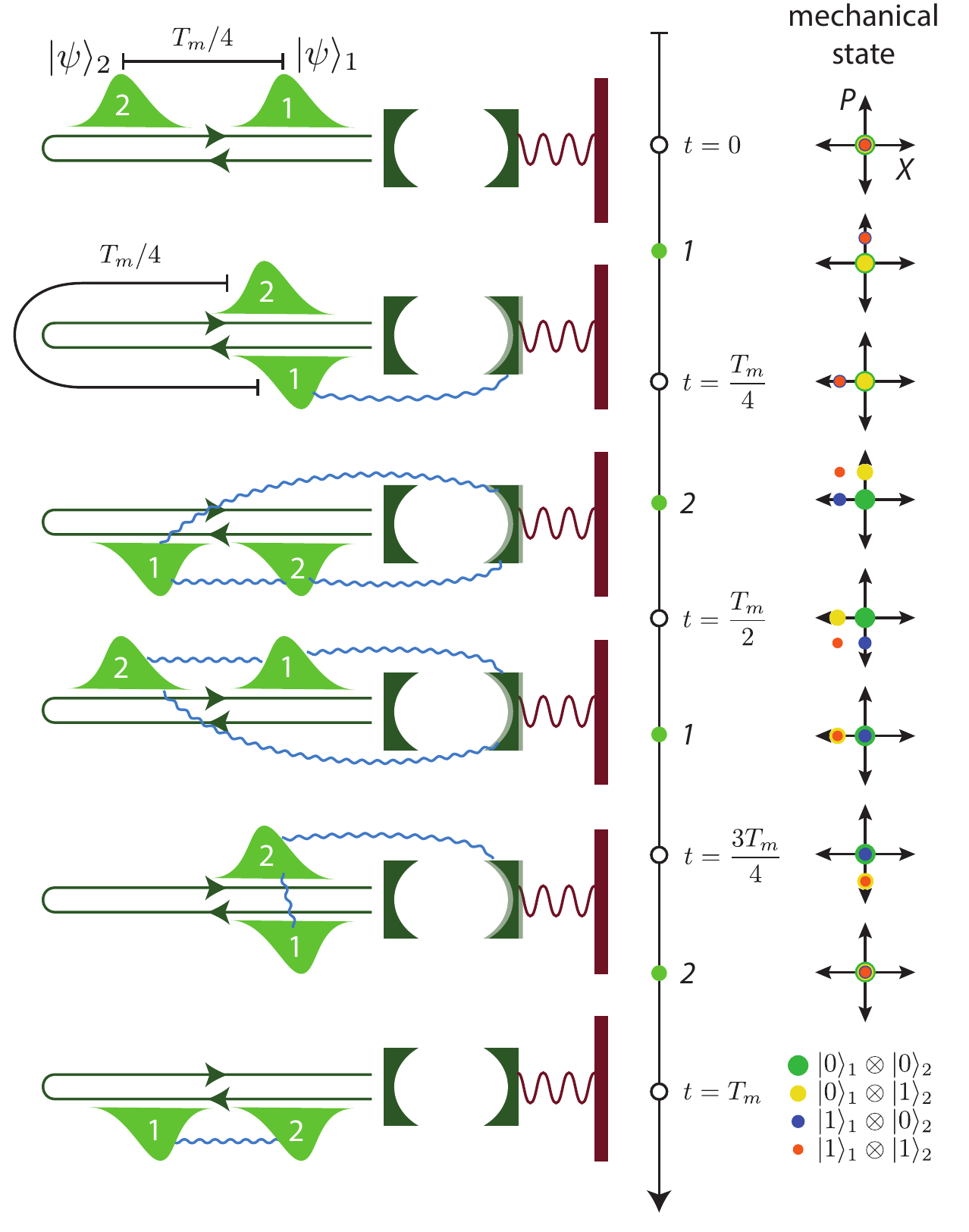}
  \caption{Protocol for the {\small{CPHASE}} gate. Two temporal photonic modes in the waveguide are separated by a time interval of $ T_m / 4 $. The modes can be in states $\ket{0}$ and $\ket{1}$. (1) The first photon mode interacts and is entangled with the mechanical oscillator since $ \ket{0} $ and $ \ket{1} $ will lead to different changes in the mechanical momentum and cause a state-dependent change in the mechanical state. After the mechanical system evolves for $T_m / 4$, (2) The second photon mode interacts with the mechanical oscillator and the mechanical system's state become entangled with both photons. After the mechanical system evolves again for another ${T_m / 4}$, (3) The first photon temporal mode interacts again with the mechanical system causing its state to be disentangled from the mechanical oscillator. After another ${T_m / 4}$ of evolution, (4) The second photon mode comes back again and the mechanical system is decoupled from both temporal modes since it goes back to its ground state regardless of the initial states of the photon modes. The blue lines signify the entanglement in the system and on the right the state-dependent evolution of mechanical state is shown in phase space.}
  \label{fig:protocol}
\end{figure}

The protocol for the {\small{CPHASE}} gate between two temporal modes of the light field is shown in Figure~\ref{fig:protocol}. The center frequencies of these two modes are chosen to be at the cavity frequency and each have a bandwidth much smaller than the cavity linewidth. They are also separated by a time $T_m/4\gg\tau$, where $T_m = 2\pi/\omegam$ is the period of the mechanical oscillation. Our system is  similar to and inspired by pulsed optomechanical experiments~\cite{Vanner2011,Vanner2013} with the distinction that no measurement occurs in our feedback network -- the photons are fed back to the system coherently. The choice of delay time is essential for disentangling the mechanical system from the photons at the end of the protocol. The mechanical system must return to its initial state, the ground state, regardless of the state of the input temporal modes, since any residual entanglement between the temporal modes and the mechanical system will reduce the fidelity of the {\small{CPHASE}} gate. As shown in Figure~\ref{fig:protocol} and is justified below, this condition is satisfied for the temporal mode spacing we have chosen.

\textbf{Semiclassical model.} Here we use a simplified semiclassical model to understand some of the behaviors of this system. The optomechanical system's Hamiltonian is given by
\begin{equation}
\op{H}{S}/\hbar = \omegao \opd{a}{} \op{a}{} + \omegam\opd{b}{} \op{b}{} + g_0 \opd{a}{} \op{a}{} (\opd{b}{} + \op{b}{}).
\end{equation}
The radiation pressure force on the optomechanical system ($\op{F}{\text{RP}} = -\hbar g_0 \aopd \aop /\xzp$) integrated over the interaction time of the photon with the cavity ($\tau \ll \omegam^{-1}$) causes a rapid change in the momentum of the mechanical system by $\Delta \op{p}{\text{RP}} = -\hbar g_0 \int_\tau \textrm{d}t \aopd(t) \aop(t)  /\xzp$. An input field $\op{c}{}(t)$ from the waveguide is incident on the cavity at time $t$. If we assume that the cavity can be eliminated adiabatically ($\kappa$ is large), the field in the cavity and waveguide may be related as $\aop(t) \approx 2\op{c}{}(t)/\sqrt{\kappa}$, so after the interaction with the propagating photon, an impulse of $\Delta \op{p}{\text{RP}} = -4\hbar g_0 \int_\tau \textrm{d}t \opd{c}{}(t) \op{c}{}(t)  /\xzp\kappa$ is imparted onto the mechanical system. We define a photon number operator $\op{n}{k} = \int_{k} \textrm{d}t \opd{c}{}(t)\op{c}{}(t)$, which counts the number of excitations in the $k^\text{th}$ temporal mode. The interaction between the $k^\text{th}$ temporal mode and the mechanical system is given by the unitary operator 
\bea
\op{U}{k} &=& \exp\left(-i\Delta\op{p}{RP}\op{x}{}/\hbar\right)\nonumber\\
    &=& \exp\left(-4i g_0 \op{n}{k} \op{x}{}  /\xzp\kappa\right),
\eea
with  position $\op{x}{} = \xzp (\opd{b}{} +\op{b}{})$. 
The mechanical free evolution operator is $ \op{U}{t} = e^{-i\omega_m b^\dagger b t}$. Our protocol can then be compactly stated as $\op{U}{\text{protocol}} \ket{\Psi}_\text{wg}\ket{0}_\text{m}$, with 
\bea
\op{U}{\text{protocol}} = \op{U}{2} \op{U}{T_m/4} \op{U}{1} \op{U}{T_m/4} \op{U}{2} \op{U}{T_m/4} \op{U}{1}.
\eea
This sequence of interactions is shown schematically in Figure~\ref{fig:protocol}. We can now calculate the result of these operations on the joint state of the optomechanical system and photonic waveguide. We take state of the photonic waveguide to be initialized as $\ket{\Psi}_\text{wg} = \ket{j}_1\ket{k}_2$, the state with $j$ photons in the first temporal mode and $k$ photons in the second temporal mode.  The mechanical mode and optical cavity are assumed to be in their ground states. Then
\bea
\op{U}{\text{protocol}} \ket{j}_1\ket{k}_2\ket{0}_\text{m}=e^{i\phi_{jk}}\ket{j}_1\ket{k}_2\ket{0}_\text{m}
\eea
where $\phi_{jk} = 0$ for $jk = 00, 01, 10$ and $\phi_{jk} = \phi_1$ for $jk=11$ with
\bea
\phi_1 \equiv 32 (g_0/\kappa)^2.
\eea
According to this simplified model, obtaining a total phase shift of $\pi$ can be accomplished either by going to a large enough coupling $g_0/\kappa$, or by running the protocol multiple times, $\Nbnc \approx \pi/\phi_1$, for a smaller $g_0/\kappa$ so that a total phase shift builds up over several bounces. This requires controlling the number of bounces, which can be accomplished for example by using high extinction Mach-Zehnders that are being developed for photonic quantum information processing. To distinguish between these two approaches, it is important to balance the losses incurred over multiple bounces with the deleterious effect of using a larger coupling and fewer bounces. For losses, we consider that in any real system, in addition to the coupling between the cavity and the waveguide, there are other intrinsic optical cavity losses characterized by a loss rate $\kappain$ leading to a probability $\kappain/\kappa$ that a photon is lost on each bounce. According to the simplified semiclassical model, since the chance of photon loss in multiple runs is proportional to $\Nbnc\kappain/\kappa$, and $\Nbnc\propto \kappa^{2}$, it is always advantageous to make $\kappa$ as small as possible. This conclusion however neglects effects that arise due to strong coupling.

One of these effects is the large optical frequency shift caused by the mechanical displacement induced by the first photon, which prevents the second photon from fully interacting with the system. This prevents a perfect erasure of the information about the photons from the mechanical oscillator and causes the residual photon-phonon entanglement. A  model described in the methods section shows that interaction of the waveguide state $\ket{1}_1\ket{1}_2$ with the optomechanical system leads to a final state $e^{i\phi} \ket{1}_1 \ket{1}_2 \ket{\beta_r}_{\text{m}}$ after one repetition of the protocol, where the mechanical system is in a coherent state $\ket{\beta_r}$ and $\beta_r \propto (g_0/\kappa)^5$. Other input states do not cause a change in the mechanical state. This residual phonon occupancy for the $\ket{11}$ input state leads to a reduction in gate fidelity on the order of $|\langle 0|\beta_r\rangle|^2=e^{-|\beta_r|^2}$ which can be made very small by going to smaller coupling and a larger number of bounces. 
Other effects caused by the finite extent of $\tau$ should also be considered carefully. For example, a photon wavepacket can obtain a non-uniform position-dependent phase shift due to frequency-dependent phase response of the cavity, causing the state of the field to no longer be in the same temporal mode. Such an effect is analyzed in more details in the method section and can be made smaller by making $\kappa\tau$ \textit{larger}. Also a mechanical system can have a finite position shift during a bounce of $\Delta x \approx \Delta p_\text{RP} \tau / m$, which causes another non-uniform phase shift on the order of $4g_0 \Delta x / \kappa \xzp = \phi_1 \omegam \tau$ for a single-photon input state in a given temporal mode. This effect can be reduced by making $\omegam\tau$ \textit{smaller}. Both of these types of imperfection cause the state of the electromagnetic field to move out of the subspace spanned by $\ket{j}_1 \ket{k}_2 \ket{0}_{\textrm{m}}$ and are difficult to capture quantitatively in an analytical form. Estimates of fidelity given these effects are provided in the methods section. However, quantitative calculations are important for understanding the realizability of the protocol given state-of-the-art experimental capabilities and therefore we turn to full quantum simulations to incorporate all these effects.

\textbf{MPS simulation for quantum dynamics.} The feedback network we are considering requires a long delay due to the slow dynamics of the mechanical oscillator. Recently methods for understanding dynamics of systems in such quantum feedback networks have been proposed~\cite{Whalen2015,Grimsmo2015,Pichler2016,Tabak2016}. In addition, we are interested in understanding the evolution of the state of the photons in the waveguide and interactions induced between them by the optomechanical system. One approach to solving the full dynamics of the waveguide and system together is to discretize the waveguide into time steps $\Delta t$ that are much smaller than any of the relevant dynamics of the system and numerically evolve what is now effectively the interaction between a 1D chain of harmonic oscillators and the system~\cite{Pichler2016}. In principle, keeping track of the state of such a 1D chain is daunting due to the exponentially large Hilbert space that scales as $O(d^{T_m / 2\Delta t})$, where $d$ is the truncated dimension of each time-bin's Fock space.  Luckily, the states of the waveguide we are considering have far less entanglement than general states in the full Hilbert space,  so efficient methods for storing and evolving the states can be utilized~\cite{Vidal2003,Vidal2004,Schon2005,Schon2007,Verstraete2010,Schollwock2011,Pichler2016}. Our full quantum simulation is based on the Matrix Product State (MPS) representation of the quantum field. MPS as well as the related time-dependent Density Matrix Renormalization Group (tDMRG) techniques have already been well-established in condensed matter physics for simulating 1D quantum many-body systems.

To implement the MPS method for the continuous quantum field of the waveguide, we discretize time in small steps $ \Delta t $ and define operator $ \op{c}{n} = \frac{1}{\sqrt{\Delta t}} \int_{t_n - \Delta t}^{t_n} \op{c}{}(t)\mathrm{d}t $ for the $ n $th time-bin, where $ t_n = n\Delta t $, $ n $ is an integer and $\op{c}{}(t)$ is the field operator in time domain. Using the commutation relation $[\op{c}{}(t), \opd{c}{} (t')] = \delta (t - t')$, it is straightforward to verify that $ [\op{c}{n}, \opd{c}{m}] = \delta_{nm} $, which means the time-bins can be interpreted as independent harmonic oscillators. The state of this quantum many body system can be represented in the canonical MPS form~\cite{Vidal2003}. A spatially distributed single photon state of the waveguide is then  $\ket{1}_f = \sum_n f_n \opd{c}{n} \ket{0\ldots0}_\text{WG} = \opd{A}{f} \ket{0\ldots0}_\text{WG}$, where $\sum_n |f_n|^2 = 1$. We consider an initial state where the optomechanical system's optical and mechanical modes are both in their ground states, and consider two temporal modes of the photon with  annihilation operators $\op{A}{1}$ and $\op{A}{2}$ separated by a time interval $T_m/4$. We identify the states $\ket{jk}\equiv\opd{A}{1}^j\opd{A}{2}^k\ket{0\ldots 0}_\text{WG}\ket{0}_\text{m}\ket{0}_\text{m}$ for $j,k=0$ or 1~\footnote{The states $\ket{jk}$ are to an extremely good approximation orthogonal given a temporal separation that is much larger than their width.}, and assume an initial state of the whole system
\bea
\ket{\Psi}_\text{i} &=&  (1 + \opd{A}{1}) (1 + \opd{A}{2}) \ket{0\ldots0}_\text{WG}\ket{0}_\text{o}\ket{0}_\text{m}/2,\label{eqn:initstate}\\
&=&\frac{1}{2}(\ket{00}+\ket{01}+\ket{10}+\ket{11})\nonumber .
\eea 
As shown in Figure~\ref{fig:si_fig_mps} in the methods section, initially the optomechanical system is at the first site of the MPS. To evolve the many-body state, we sequentially update the MPS by applying the unitary $\op{U}{n} = \exp \left( -i \op{H}{S} \Delta t + \sqrt{\kappa \Delta t} (\op{a}{} \opd{c}{n} - \opd{a}{} \op{c}{n}) \right)$ as a local gate on the $ n $th time-bin and the optomechanical system.  Then a swap gate is used to permute the order of them so that $ U_{n+1} $ can be applied in the same way as $ U_n $. The swap gate is used because it is more convenient to update the MPS in a local way by applying the gates only to the nearest neighbor sites -- it does not represent a physical evolution of the many-body state. This process effectively simulates a discrete representation of the Quantum Stochastic Schr\"odinger Equation (QSSE)~\cite{Pichler2016} and is continued until the optomechanical system reaches the last site of the MPS.

In contrast to Ref.~\cite{Pichler2016} where the number of sites in the MPS is proportional to the total simulation time, here our waveguide is modeled as a finite number of time bins corresponding to the feedback waveguide length. After every $ T_m/2 $ time interval, corresponding to the total optical path length for a round trip, the optomechanical system has interacted with each time bin and reaches the last site of the MPS. In the next time step, due to the reflection off the far mirror, the system interacts with the first time bin again. This is accomplished by moving the system back to the first site of MPS via a series of swap gates. At this point, the evolution can be continued as before to simulate interaction after one bounce. This process is repeated $ 2\Nbnc $ times for $\Nbnc$ runs of the protocol.

\begin{figure}
  \centering
  \includegraphics[width=0.45\textwidth]{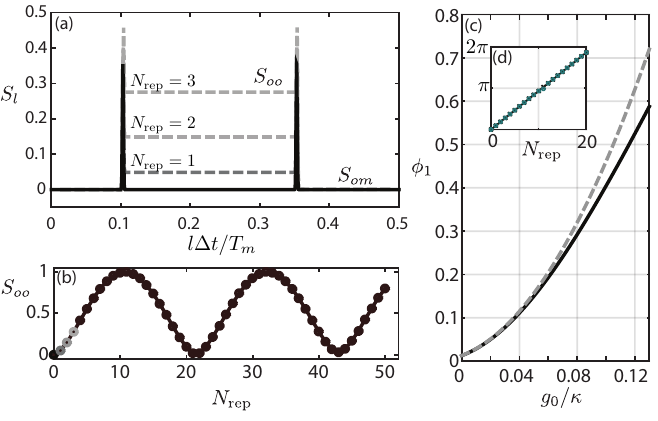}
  \caption{(a) Entanglement entropy $S_l$ of the fully coupled waveguide-optomechanical system for $\Nbnc=0,1,2,3$. (b) Evolution of the entanglement entropy $S_{oo}$ between the two localized photons over many repetitions of the protocol. The first four points are the same as those in (a). (c) Simulated phase shift $\text{Arg}(\avg{11|\Psi})$ after $\Nbnc=1$. The dashed line is comparison with the semiclassical $\phi_1$. The MPS simulations in (a) and (b) were done for $g_0/\kappa = 0.1$, $\omegam/\kappa  = 2.5\times10^{-4}$, while in (c) $g_0/\kappa$ was varied. (d) The simulated phase shift over multiple repetitions of the protocol for $g_0/\kappa = 0.1$ is plotted. The circles use $S_{oo}$ to infer $\phi$, while the black trace is calculated directly by taking the argument of $\avg{11|\Psi}$. There is good agreement between the two and the phase increases linearly as $\Nbnc\phi_1$. }
  \label{fig:result1}
\end{figure}

\textbf{Entanglement entropy and phase calculation.} MPS provides a convenient way to extract information about entanglement entropy of a quantum system. In the MPS representation we decompose the state vector of the $N$-site system, $c_{i_1 \ldots i_N}$, into a product of tensors $\Gamma^{[k]i_N}_{\alpha_{k-1}\alpha_{k}}$ and vectors $\lambda^{[k]}_{\alpha_k}$ for $k=1\ldots N$, such that a bipartition of the state at bond $l$ can be written as a Schmidt decomposition $\ket{\Psi} = \sum_{\alpha_l} \lambda^{[l]}_{\alpha_l} \ket{\Phi^{[1\ldots l]}_{\alpha_l}}\ket{\Phi^{[(l+1)\ldots N]}_{\alpha_l}}$~\cite{Vidal2004}. This allows us to calculate the entanglement entropy between the two halves of the system by simply reading off the values of the $\lambda$ vector at bond $l$, and calculating $S_l = -\sum_{\alpha_l} \lambda^{[l]2}_{\alpha_l} \log_2 \lambda^{[l]2}_{\alpha_l}$. The black line in Figure~\ref{fig:result1}(a) is a plot of $S_l$ as a function of $l \Delta t / T_m$ for the initial state $\ket{\Psi}_\text{i}$, which is essentially the bipartition entanglement entropy throughout the whole waveguide. The two peaks at the positions of the two photons correspond to the finite width photon excitations in the waveguide, since one excitation is spread across multiple local sites and detecting the photon at a particular bin tells us that there are no photons in the nearby bins. Stated more concisely, $\ket{1}_f$ cannot be written as a product state of time-bin localized photon excitations. In addition to the localized peaks, two other interesting regions are that between the photons, and the region between the photons and the optomechanical system. We call the entanglement entropy between the photons and mechanical resonator $S_{om}$ and the entanglement entropy between the two photons $S_{oo}$. When calculating $S_{oo}$ we are in fact finding the entanglement between one temporal mode of the waveguide and the rest of the system, including the other temporal mode as well as the optomechanical system. However, when $S_{om}$ is close to 0, \ie, mechanical system disentangles from photons, $S_{oo}$ simply becomes the entanglement between two photon temporal modes. As expected, both $S_{om}$ and $S_{oo}$ are zero since the initial state can be written as a product state of separated temporal  modes of the optical field in the waveguide and the optomechanical system, c.f. equation~(\ref{eqn:initstate}).

The entanglement entropy throughout the waveguide is plotted after a few subsequent runs of the protocol and shown in dashed lines in Figure~\ref{fig:result1}(a). For the system parameters in Figure~\ref{fig:result1}(a), $S_{oo}$ increases on each run of the protocol while $S_{om}$ remains close to zero. The evolution of $S_{oo}$ for more runs of the protocol is plotted in Fig.~\ref{fig:result1}(b) and shows clear oscillatory behavior. This can be attributed to the evolving phase shift of the $\ket{11}$ state with respect to the other states, since for a state $ \ket{\psi} = \frac{1}{2} (\ket{00} + \ket{10} + \ket{01} + e^{i\phi} \ket{11}) $ the entanglement entropy is $S = 1 - [(1+c)\log_2(1+c) + (1-c)\log_2(1-c)]/2$, where $c = \cos (\phi/2)$. To verify that this is in fact the correct interpretation, we calculate the phase shift directly from the wavefunctions by taking the overlap between a vector $\ket{11} = \opd{A}{1}\opd{A}{2}\ket{0\ldots 0}_\text{WG}\ket{0}_\text{m}\ket{0}_\text{m}$ evolved on a system with $g_0 = 0$, and the result of our simulation with non-zero $g_0$ after each run of the protocol. The reason for doing this instead of comparing with the initial unevolved state is that the temporal mode of the photon indeed changes but the distortion caused by a cavity can in principle be reversed using other linear optical components~\cite{Marshall2003}. The total phase shift $ \phi $ increases linearly as $ \Nbnc\phi_1 $ in Fig.~\ref{fig:result1}(d), which explains the oscillatory behavior of entanglement entropy (in fact these phase shift gives exactly the same entanglement entropy as Fig.~\ref{fig:result1}(b)). We compare the phase shift after one run of the protocol to the semiclassically predicted results $\phi_1 = 32(g_0/\kappa)^2$, and find good agreement at smaller $g_0/\kappa$ as shown in Figure~\ref{fig:result1}(c). The agreement becomes progressively worse at higher coupling. Finally, we calculate in the same way the phase shift for $\ket{00}, \ket{10}, \ket{01}$ components and find them all 0 for the values of $g_0/\kappa$ in Figure~\ref{fig:result1}.

\begin{figure}
  \centering
  \includegraphics[width=0.5\textwidth]{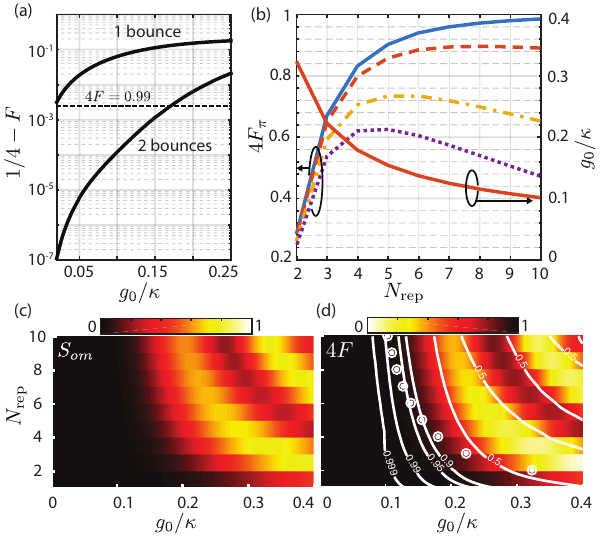}
  \caption{(a) Reduction of modal overlap $F=|\avg{11|\Psi}|^2$ as a function of coupling strength $g_0/\kappa$. The infidelity is very low after two bounces and increases with coupling strength. After only one bounce, the infidelity is much larger, since the mechanical system is no longer in its ground state. (b) The fidelity $4F_\pi$ is plotted along with the required $g_0/\kappa$ to obtain a $\pi$ phase gate for each $\Nbnc$. The fidelity improves monotonically with increasing $\Nbnc$ and with small $g_0/\kappa$. Taking into account losses per repitition $\eta$ ($\eta = 0.99$ dashed, $\eta = 0.96$ dot-dashed, and $\eta=0.93$ dotted), a maximum in $4F_\pi$ is observed at different values of $\Nbnc$. (c) and (d) The residual photon-phonon entanglement $S_{om}$ (c) as well as the fidelity of the {\small{CPHASE}} gate (d) for different values of $ g_0/\kappa$ and $\Nbnc$. The white points in (d) represents values of $\Nbnc$ and $g_0/\kappa$ that implement a $\pi$ phase gate, which are the data points for (b). Here $\omegam/\kappa = 1.5\times 10^{-4}$ and the temporal width of the photon $\tau = 1000/\kappa$.}
  \label{fig:result2}
\end{figure}

\textbf{Fidelity of the {\small{CPHASE}} gate.} In addition to calculating the phase, we  verify that the state of the electromagnetic field after interaction with the optomechanical system remains within the subspace of states $\ket{jk}$, modulo the linear optical response of cavity. We calculate the fidelity $F = |\langle 11|\Psi\rangle|^2$ after one and two bounces of the photons from the cavity. Nominally, after two bounces, \ie, $\Nbnc=1$, we  expect $F=1/4$ for the input state in equation (\ref{eqn:initstate}). In Figure~\ref{fig:result2}(a) a plot of the infidelity $1/4 -F$ against the interaction rate is shown for one and two bounces of the waveguide photons from the cavity. After one bounce, the significant entanglement between the mechanical system and waveguide photons leads to higher infidelity. After two bounces, \ie, a full run of the protocol, the entanglement between the mechanics and waveguide photons is largely erased, causing an increase in the overlap $F$. Larger interactions cause a breakdown of this picture and lead to residual entanglement with the mechanical system and lower fidelity. For $g_0/\kappa=0.1$, we simulate a fidelity of $4F \approx 0.9996$ after two bounces, and $4F \approx 99\%$ after the $\Nbnc=10$. These simulations show that the photon wavefunction is not significantly distorted in the interaction, as would be expected for large instantaneous Kerr nonlinearities~\cite{Shapiro2006a}.

To obtain a $\pi$ phase gate, multiple bounces are required. In Fig.~\ref{fig:result2}(b), we show the simulated fidelity, $F_\pi = |\langle 11|\Psi\rangle|^2$, and the required $g_0/\kappa$ for phase gates implemented with different numbers of $\Nbnc$. For phase gates with larger $\Nbnc$, smaller $g_0/\kappa$ are required; the gate fidelity is found to increase monotonically as the system becomes better described by the semiclassical model. Photon loss causes a reduction in the fidelity by a factor $\eta^{\Nbnc}$, where $\eta$ is the probability of a photon being lost on a single repetition of the protocol. This finite photon loss causes the gate fidelity to be maximized at a finite $\Nbnc$, with smaller $\Nbnc$ being favored for higher losses $\eta$. 

Another signature of failure of the {\small{CPHASE}} gate is the presence of residual entanglement between the photons and the mechanical system after running the protocol, \ie, a non-zero value for $S_{om}$. In Fig.~\ref{fig:result2}(c) and (d), the evolution of $S_{om}$ and gate fidelity are plotted against $g_0/\kappa$ and $\Nbnc$. It is clear that increasing $g_0/\kappa$ causes an increase in the residual entanglement and reduction in gate fidelity. The white points in Fig.~\ref{fig:result2}(d) outline the relationship between $\Nbnc$ and $g_0/\kappa$ needed to obtain a phase shift of $\pi$ according to the quantum simulations. From these two figures, it is clear that  the reduction in fidelity is largely due to residual photon-phonon entanglement, which can be approximated semiclassically. In the methods section, we outline how the semiclassical model predicts a residual phonon occupancy giving qualitatively similar results to the quantum case, though the full quantum calculations are more forgiving in terms of obtainable phase shifts. 

By studying the simulated fidelity for different parameters, we can make the phase gate conditions more precise. For $\Nbnc=4$, we find that the bounds $\omegam\tau<0.3$ and $\kappa\tau>200$ are sufficient to prevent excess loss in fidelity. In our simulations the temporal width of the phonons $\tau$ was chosen to be $\tau=0.15/\omegam=1000/\kappa$.

\section*{Discussion}
\textbf{Dissipation and Experimental Prospects.} We have shown that strong photon-photon interactions can be obtained with an optomechanical system that is outside the strong coupling regime ($g_0\kappa/\omegam^2 > 1$). We estimate the effect of five sources of decoherence; (1) the mechanical coupling to the thermal bath, (2) the effect of other weakly-coupled mechanical modes and sources of phase fluctuations, (3) the intrinsic optical loss $\kappain$, (4) insertion loss, and (5) the propagation losses in the long waveguide. The important parameter in understanding the mechanical decoherence (1) is the thermalization rate $\Gamma_\text{m} = n_b\omegam/Q_\text{m}$, where $\omegam/Q_\text{m}$ is the mechanical linewidth and $n_b$ is the thermal occupation at the mechanical frequency. In the high temperature limit, $n_b = kT/\hbar\omegam$ and $\Gamma_\text{m} = kT /\hbar Q_\text{m}$. As long as the $\Gamma_\text{m}\times \Nbnc T_m \ll 1$, the chance of a phonon entering the system from the bath during the time that protocol is being run for remains small. This is equivalent to having $f_\text{m} Q_\text{m} \gg \Nbnc kT/\hbar$. Such $f_\text{m} Q_\text{m}$ products have been obtained at cryogenic tempeatures~\cite{Chan2011}, and more recently at room temperature with silicon nitride membranes~\cite{Norte2016,Reinhardt2016,Tsaturyan2016}. The effect of competing mechanical modes (2) has long been an issue in pulsed optomechanics experiments~\cite{Vanner2011}. Recent experiments~\cite{Vanner2013} have been successful at reducing the coupling to these modes to a few percent of the primary mode by careful engineering of structures and positioning of the optical beam. In our case weak coupling to parasitic mechanical modes leads to dephasing and is considered more carefully in the methods section. Sources (3)-(5) affect the protocol in a similar way, in the sense that they introduce a finite probability $\eta$ that a photon can be lost to the environment during the execution of the gate. For the intrinsic losses in the cavity (3), this means requiring $\eta_i^{2\Nbnc} \approx (1 - \kappain/\kappa)^{2\Nbnc} \approx 1 - 4\Nbnc\kappain/\kappa$~\footnote{We assume throughout this discussion that total loss rate $\kappa$ is composed of the intrinsic and extrinsic parts $\kappaex +\kappain$, with $\kappaex \gg \kappain$.} be close to $1$. This linear reduction in fidelity is seen in Fig.~\ref{fig:result2}b and makes smaller $\Nbnc$ more optimal. Insertion losses (4) can be considered in an identical way. Finally (5), the chance of a photon being absorbed in the long delay is related to the attenuation length of the fiber (on the order of $0.15~\text{dB/km}$ in the telecom C-band) and the required propagation path needed in the fiber, $\Nbnc T_m v_\text{fiber}$. To achieve a more than $90\%$ chance for the photon to survive propagation in the delay for the $\Nbnc=4$ case, we require a mechanical frequency greater than $260~\text{kHz}$. Together these parameters mean that $\Nbnc=4$ and $(g_0,\omegam,\kappa)/2\pi=(31.3~\text{MHz},260~\text{kHz},173~\text{MHz})$ are needed to implement the CPHASE gate with $F \approx 80\%$. Currently, these parameters, though significantly easier to obtain than the large $g_0/\kappa$ and good cavity limit $\kappa<\omegam$, remain outside of our experimental reach. Nonetheless, progress in experimental quantum optomechanics, and techniques for enhancing coupling by stacking low frequency membranes~\cite{Xuereb2012}, are expected to bring us to the required parameters in the coming years.

In conclusion, we have demonstrated that a {\small{CPHASE}} gate between propagating photons can be implemented with the slow and weak nonlinearity of an optomechanical resonator under the correct quantum feedback conditions. In addition to opening up a new part of the optomechanical parameter space to quantum experiments, the approach shows that quantum feedback has the potential to enhance quantum nonlinear phenomena. In the future, it is interesting to consider the evolution of entanglement in quantum feedback networks using the MPS method, to better understand how interesting quantum many-body states with nontrivial correlations can be generated.

\textbf{Acknowledgements.} This work was supported by NSF ECCS-1509107 and the Stanford Terman Fellowship, Tsinghua University undergraduate research program, ONR MURI QOMAND, as well as start-up funds from Stanford University. We thank Jeff Hill and Marek Pechal for their assistance.

\section*{Methods}
Here we will go through some details of the matrix product state method,  giving an explicit algorithm to decompose a single photon state with wide temporal extent into MPS form and also showing the equivalence between our simulation scheme and the dynamics of a closed waveguide system. We will also present results of fidelity simulations not included in the main text as well as a more detailed semiclassical analysis and the effect of parasitic mechanical modes.

\textbf{Time-bin representation.} The system-bath Hamiltonian $ \op{H}{} = \op{H}{S} + \op{H}{B}+ \op{H}{\mathrm{int}} $, where
\begin{equation}
\begin{split}
\op{H}{S} &= \omega_m \opd{b}{}\op{b}{} + g_0 \opd{a}{}\op{a}{} (\opd{b}{}+\op{b}{}) \\
\op{H}{B} &= \int_{-\infty}^{\infty} \mathrm{d}\omega \omega \opd{c}{}(\omega)\op{c}{}(\omega) \\
\op{H}{\mathrm{int}} &= i\int_{-\infty}^{\infty} \mathrm{d}\omega \sqrt{\frac{\kappa}{2\pi}} (\op{a}{} \opd{c}{}(\omega) - \opd{a}{} \op{c}{}(\omega)) .
\end{split}
\end{equation}
The operators $\op{a}{}$, $\op{b}{}$ and $\op{c}{}(\omega)$ are annihilation operators for photons in the cavity, phonons of mechanical system, and photons in the waveguide with mode index $\omega$. 
Note that in writing this Hamiltonian, we are already in the rotating frame of the optical cavity frequency $ \omegao $. We can further go  into the rotating frame with respect to the bath Hamiltonian $ \op{H}{B} $ and the interaction term becomes
\begin{equation}
\begin{split}
\op{H}{\mathrm{int}} (t) &= i\int_{-\infty}^{\infty} \mathrm{d}\omega \sqrt{\frac{\kappa}{2\pi}} (\op{a}{} \opd{c}{}(\omega) e^{i\omega t} - \opd{a}{} \op{c}{}(\omega) e^{-i\omega t} ) \\
&= i \sqrt{\kappa} (\op{a}{} \opd{c}{}(t) - \opd{a}{} \op{c}{}(t) ) ,
\end{split}
\end{equation}
where we have defined
\begin{equation}
\op{c}{}(t) = \frac{1}{\sqrt{2\pi}} \int_{-\infty}^{\infty} \mathrm{d}\omega \op{c}{}(\omega) e^{-i\omega t} .
\end{equation}

To calculate the evolution of the whole system, it is convenient to discretize time in small steps of $ \Delta t $. Define the quantum noise increments in the time domain for the input field as:
\begin{equation}
\op{c}{n} = \frac{1}{\sqrt{\Delta t}} \int_{t_n - \Delta t}^{t_n} \op{c}{}(t)\mathrm{d}t,
\end{equation}
where $ t_n=n\Delta t $ and $ n $ is an integer. It is straightforward to verify that $ [\op{c}{n}, \opd{c}{m}] = \delta_{nm} $ from the commutation relation $ [\op{c}{}(\omega), \opd{c}{} (\omega ')] = \delta (\omega - \omega^\prime) $ of the field mode, which means all the time-bins can be interpreted as independent harmonic oscillators. Thus the Hilbert space for the whole system is a tensor product space, including contributions from optomechanical system and quantum field in the waveguide, which can be modeled as a series of harmonic oscillators.

In the time-bin representation, the evolution of the system state in the $ n $th time step is
\begin{equation}
\begin{split}
\ket{\Psi(t_{n+1})} &= \op{U}{n} \ket{\Psi(t_n)} \\
&= \exp \left( -i \op{H}{S} \Delta t + \sqrt{\kappa \Delta t} (\op{a}{} \opd{c}{n} - \opd{a}{} \op{c}{n}) \right)\ket{\Psi(t_n)}.
\end{split}
\end{equation}

\textbf{Initial state preparation.} As usual, at $ t=0 $ the state of the optomechanical system and the quantum field in the waveguide are assumed to be completely uncorrelated and separable. The state of the waveguide is composed of two temporal modes that are spaced apart from each other. A large separation in time means that the state of the quantum field is very nearly the product state of two single photon states in two independent temporal modes.  A single photon in a temporal mode has a wavefunction given by $\ket{1}_f = \int f(t) \text{d}t \op{c}{}(t) \ket{\text{vac}}_\text{WG} = \opd{A}{f} \ket{\text{vac}}_\text{WG}$. This can be rewritten in the time-bin representation as 
\bea
\sum_n f_n \opd{c}{n} \ket{0\ldots0}_\text{WG} = \opd{A}{f} \ket{0\ldots0}_\text{WG}.
\eea
More generally, the joint state of the waveguide with two temporal modes each in their 0 or 1 state can be described by the linear combination of state vectors $\ket{jk}\equiv\opd{A}{1}^j\opd{A}{2}^k\ket{0\ldots 0}_\text{WG}\ket{0}_\text{m}\ket{0}_\text{m}$ for $j,k=0$ or 1. Specifically, in our simulation the initial state is choosen to be
$ \ket{\Psi}_\text{i} = \frac{1}{2}(\ket{00}+\ket{01}+\ket{10}+\ket{11}) $, as shown in the main text.

\textbf{Matrix product state.} For a general many body state
\bea
\ket{\Psi} = \sum_{\{i\}} c_{i_1, \cdots, i_n} \ket{i_1} \otimes \cdots \otimes \ket{i_n}
\eea
the following canonical decomposition always exists~\cite{Vidal2003}:
\bea
c_{i_1, \cdots, i_n} = \Gamma_{\alpha_1}^{[1]i_1} \lambda_{\alpha_1}^{[1]} \Gamma_{\alpha_1 \alpha_2}^{[2]i_2} \lambda_{\alpha_2}^{[2]} \Gamma_{\alpha_2 \alpha_3}^{[3]i_3} \cdots \Gamma_{\alpha_n}^{[n]i_n}
\eea
where the state is represented using a series of tensors $ \Gamma $ and vectors $ \lambda $. An important property of this canonical form is that for any $ k $, $ \lambda^{[k]} $ is the Schmidt vector for a certain bipartition of the whole system into subsystem $ 1 \rightarrow k $ and subsystem $ k+1 \rightarrow n $. For a quantum state with restricted amount of entanglement, the Schmidt vectors can be truncated by some threshold ($10^{-4}-10^{-5}$ in our case -- verifying that our results are not dependent on this threshold), which enables efficient classical simulation.

\textbf{Single photon state decomposition.} Generally speaking, two things are required for the MPS simulation: preparing the many-body state $ \ket{\Psi(t)} $ in MPS form and updating this state step by step to evolve $ \ket{\Psi(t)} $. In this and the next section, we consider these two aspects.

To decompose the initial state in an MPS form, a sequence of singular value decompositions (SVD) are required. Note that for separated temporal modes each being in the 0 or 1 photon states, the dimension of Hilbert space for any of the time-bins can be chosen to be 2. In addition, we are operating in the polynomial $O(N^2)$ subspace of the $O(d^N)$ dimensional Hilbert space of the waveguide --  only states $ \ket{11\cdots 0}, \ket{101\cdots 0}, \cdots \ket{011\cdots 0}, \cdots, \ket{0\cdots 11} $ are used since we ignore the possibility of two photons localizing at the same time-bin. Considering $M$ to be the dimension of the truncated Hilbert space for the optomechanical system, the state of the full system would require $O(N^2M)$ parameters. The MPS representation takes advantage of this reduction in problem size in an implicit way. In addition, we note that  though we are limited  to this subspace due to the photon number conserving symmetry of the Hamiltonian and our selection of the initial state, our implementation of the MPS method is capable of simulating the dynamics for a more complex input state in an efficient and automatic way, since the Hilbert space is never explicitly reduced.

The initial state decomposition can be further simplified using the fact that it is a product state of two single photons localizing at different parts of the waveguide. Therefore the problem is reduced to decompose a single photon state for that part of the waveguide. Imagine a single photon is localized at $n$ adjacent time bins with the state $\sum_{i=1}^{n} f_i \opd{c}{i} \ket{0\ldots0}_\text{n}$. The decomposition algorithm is as follows:
\begin{enumerate}
\item For the first step, calculate the density matrix for the first time-bin and diagonalize it to get its eigenvalues $ \lambda^{[1]2}_{\alpha_1} $ and eigenvectors $ \ket{\Psi_{\alpha_1}^{[1]}} $ where $\alpha_1$ is the label of different eigenvalues. The Schmidt vector $ \lambda^{[1]} $ is the square root of the eigenvalues, which can be truncated by keeping all those values of $\alpha_1$ such that $\lambda^{[1]}_{\alpha_1}$ is larger than a certain threshold. The elements of tensor $ \Gamma^{[1]} $ can be obtained by expanding the eigenvectors in the local basis \{$ \ket{0}, \ket{1} $\}, \ie, $ \langle i_1\ket{\Psi_{\alpha_1}^{[1]}} = \Gamma^{[1]i_1}_{\alpha_1} $. Here $i_1$ can take the value of 0 and 1 and $\alpha_1$ only takes the values after truncation.
\item For the $ k $th ($ 1<k \leq n $) step, take the composite system of $ 1 \rightarrow k $ time-bins as a whole and calculate its density matrix. Similarly, calculate the square root of the eigenvalues and then truncate to get the Schmidt vector $ \lambda^{[k]} $. Take the inner product $ \bra{\Psi_{\alpha_{k-1}}^{[k-1]}} \langle i_k\ket{\Psi_{\alpha_k}^{[k]}} $ to give the element $ \Gamma^{[k]i_k}_{\alpha_{k-1}\alpha_{k}} $ of $ \Gamma^{[k]} $, where $ \ket{i_k} $ is the local basis \{$ \ket{0}, \ket{1} $\} of the $ k $th time-bin and $\alpha_{k-1}, \alpha_k$ are the index of the truncated Schmidt vectors $\lambda^{[k-1]}$ and $\lambda^{[k]}$ respectively.
\end{enumerate}

\textbf{MPS update.} To evolve the many body state, a sequence of unitary operators $\op{U}{n}$  are applied to update the MPS. Here
\begin{equation}
\op{U}{n} = \exp \left( -i \op{H}{S} \Delta t + \sqrt{\kappa \Delta t} (\op{a}{} \opd{c}{n} - \opd{a}{} \op{c}{n}) \right).
\end{equation}
As already mentioned in the main text, the MPS used for simulation of our protocol keeps track of a waveguide of finite length equivalent to a delay time of $ T_m/2 $ and also the optomechanical system.
\begin{figure}
  \centering
  \includegraphics[width=0.45\textwidth]{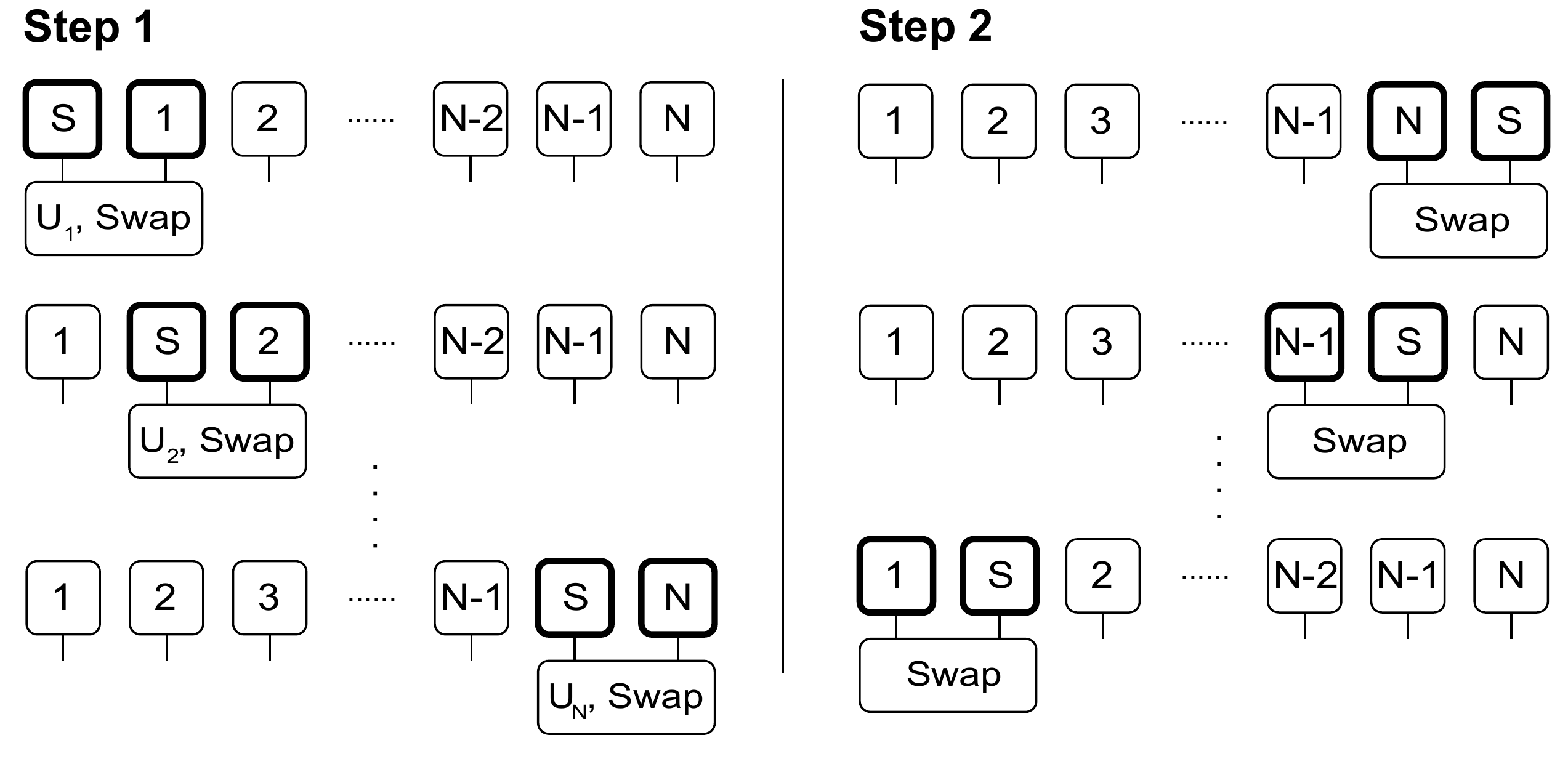}
  \caption{Two steps for MPS update. Step 1: From 1 to $N$, Repeatedly apply gate $ \op{U}{n} $ to the optomechanical system and the $ n $th time-bin, followed by a swap gate to permute the order of them. Step 2: Use a series of swap gates to bring the optomechanical system back to the begining of the MPS chain from the end, so that steps 1 and 2 can be repeated again to finish one repetition of the protocol.}
  \label{fig:si_fig_mps}
\end{figure}

The update of the MPS includes the following steps (here $ N $ is the total number of time-bins) as illustrated in Figure~\ref{fig:si_fig_mps}:
\begin{enumerate}
\item For $ n $  from 1 to $ N $, apply $ \op{U}{n} $ to the optomechanical system and the $ n $th time-bin. Then use a swap gate to permute the order of them, so that $ \op{U}{n+1} $ can be applied in the same local way as $ U_n $. By the end of this step, the optomechanical system is at the last site of the MPS.
\item For $ n $  from $ N $ to 1, perform swap gate to the optomechanical system and the $ n $th time-bin. Note that in this step the many-body state is not evolved in time since it stays the same under swap gates.
\item Repeat steps 1 and 2 to complete one repetition of the protocol. The full evolution may be over multiple repetitions, $\Nbnc$.
\end{enumerate}

\textbf{Extraction of the entanglement entropy.} The canonical MPS form provides a convenient way to extract many properties of the system, like average values during evolution, correlation functions of the quantum field and so on. In addition to the phase and fidelity that can be directly calculated from the wavefunction, we are particularly interested in the entanglement entropy of various bipartitions of the whole system. These can be directly calculated from the $ \lambda^{[k]} $, the Schmidt vector at time $t_k$, in the MPS representation. We calculate
\begin{equation}
S_k = -\sum_{\alpha} \lambda^{[k]2}_\alpha \log_2 \lambda^{[k]2}_\alpha ,
\end{equation}
which gives us the entanglement  between part of the system, containing time-bins $ 0<t<t_k $, and another part, with time-bins $ t_k<t<T_m/2 $.

\textbf{Relation to a closed system.} To simulate the effect of many reflections by the mirror, we use a series of swap gates to move the optomechanical system back every time it reaches the last site of the MPS. This scheme seems reasonable, and we justify it here in a more rigorous way. The  system as modelled is simply a closed system without losses. Here we wish to show formally the equivalence between our simulation scheme and the dynamics of the closed system.

Consider a long waveguide of length $ L $ with linear dispersion, then $ \op{H}{\text{B}} = \omega_0 \sum_{-\infty}^{\infty} n \opd{c}{n}  \op{c}{n} $ where $ \omega_0 = \pi c/L $ is the frequency of the fundamental mode and $ c $ is the speed of light. The summation starts from $ -\infty $ since we are in a rotating frame of cavity frequency. In this model the delay time of the coherent feedback is $ T = 2L/c $.

After going into the rotating frame of $ \op{H}{\text{B}} $, the interaction term is
\begin{equation}
\begin{split}
\op{H}{\mathrm{int}} &= \sqrt{\frac{\kappa}{2\pi}} \sum_{n} (\op{a}{} \opd{c}{n}e^{in\omega_0 t} - \opd{a}{} \op{c}{n} e^{-in\omega_0 t}) \\
&= \sqrt{\kappa} (\op{a}{} \opd{c}{} (t) - \opd{a}{} \op{c}{} (t) ) ,
\end{split}
\end{equation}
where we have defined $ \op{c}{}(t) = \frac{1}{\sqrt{2\pi}} \sum_n \op{c}{n} e^{-in\omega_0 t} $. It is important to note that $ \op{c}{}(t) = \op{c}{}(t+2\pi/\omega_0) = \op{c}{}(t+T) $ which is a direct result of the coherent feedback.

Next let us discretize time in small steps of $ \Delta t = T/N $, where $ N $ is a large integer and also the number of time-bins. We define in the time domain $ \op{c}{i} = \frac{1}{\sqrt{\Delta t}} \int_{t_i - \Delta t}^{t_i} \op{c}{}(s)\mathrm{d}s $, where $ i $ is an integer and $ t_i = i\Delta t $. Since $\op{c}{i+N} = \op{c}{i} $ for any $ i $, all the independent operators are $ \{ \op{c}{i}, i = 1, \cdots, N \} $. As before, it is straightforward to show that $ [\op{c}{i},\opd{c}{j}] = \delta_{ij}, \forall 1\leqslant i, j \leqslant N $. Therefore we have $ N $ independent harmonic oscillators as time-bins.

To describe the evolution of this state, we start again with the single-step evolution
\begin{equation}
\op{U}{n} = \exp \left( -i \op{H}{S} \Delta t + \sqrt{\kappa \Delta t} (\op{a}{} \opd{c}{n} - \opd{a}{} \op{c}{n}) \right).
\end{equation}
Here $ \op{c}{n} $ is the operator defined in the time domain and $\op{U}{n} = \op{U}{n+N}$. This cyclic property of these unitary matrices evolving the state means that the total evolution operator $\prod_j \op{U}{j}$ over many steps can be written as
\begin{equation}
\ket{\Psi(t)} = \op{U}{m} \cdots \op{U}{1} (\op{U}{N}\cdots \op{U}{1})^n \ket{\Psi(0)}
\end{equation}
for $ t = nT+m\Delta t $ where $ m,n $ are two integers and $ 1 \leqslant m \leqslant N $. When $ m $ goes from 1 to $ N $, the optomechanical system moves through the MPS and interacts with the photons. When $ m $ becomes 1 again, a series of swap gates are required to move the optomechanical system back to the first site of the MPS, so that $ \op{U}{1}$ can be applied on the MPS in a local way. This is exactly the simulation scheme we have implemented as described above.

\textbf{Simulation Parameters.} The cutoff of the singular values in the singular value decomposition calculation is set to be $10^{-4}$. Since the coupling between the cavity and the waveguide is the fastest process in the rotating frame, we choose $ \Delta t = 1/(2\kappa) $. The temporal width of the photon is chosen to be $\tau =  1000/\kappa $, \ie, most of the photon is spread across $ 1000\times 2=2000 $ time-bins to ensure that it can fully enter the cavity. We choose the cutoff of the Fock space to be $2$ for the optical cavity, $15$ for the mechanical oscillator and $2$ for each time bin. The total waveguide is composed of $T_m/(2\Delta t)$ time-bins, which is $ 41,888 $ for $\omegam/\kappa = 1.5\times 10^{-4}$. We have verified that these paramters are sufficient for capturing the physics of the problem.

\begin{figure}
  \centering
  \includegraphics[width=0.45\textwidth]{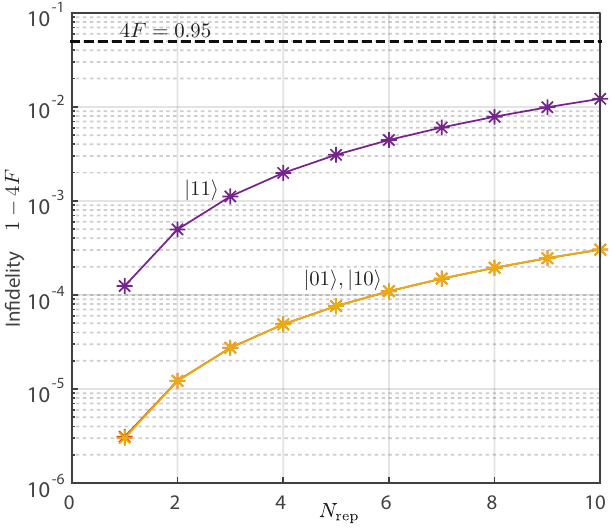}
  \caption{Plot of infidelity $(1-4F)$ of the resulting state of the waveguide. These are results of simulations done for $g_0/\kappa = 0.1$, $\omegam/\kappa  = 1.5\times10^{-4}$, $\tau = 1000/\kappa$, and $\Nbnc=1,\ldots,10$. Note that for $\ket{00}$ subspace the results are always 0 and therefore not shown here.}
  \label{fig:si_fig}
\end{figure}

\textbf{Evolution of Fidelity over many bounces.} The quantum optomechanical nonlinearity can cause the state of the quantum field in the waveguide to move outside of the subspace defined by the four state vectors $\ket{jk}$, where there are $j,k=0,1$ photons in the temporal modes $j$ and $k$ of the waveguide. There are three effects that cause this. One is the change in the temporal mode after each bounce from the cavity. The other is leakage due to the nonlinearity. Finally there is the effect of residual entanglement with the mechanical system after a repetition of the protocol. We are interested in the latter two since the first can be in principle reversed with linear optical components. We calculate  the change in the overlaps $|\avg{jk|\Psi}|^2$ starting from the initial state $\ket{\Psi}_\text{i}$ as defined in the main text, and with $\ket{jk}$  evolved separately with $g_0 = 0$. The results are plotted in Figure~\ref{fig:si_fig}. Nominally, $F=0.25$ in the perfect case for every $j$ and $k$, and so the infidelity is $1-4F$. 

\textbf{Semiclassical calculation of fidelity.} The fidelity of the \small{CPHASE} gate depends on the input state. In a semiclassical model in the bad cavity regime and with photons with sufficient temporal extent $\tau$, the fidelity of the gate for $\ket{00}$, $\ket{10}$ and $\ket{01}$ input states is close to one. However, for $\ket{11}$ input state, the  cavity frequency shift due to  the interaction of the first photon reduces the probability of the second photon entering the cavity and causes a residual phonon number at the end of the protocol.

\begin{figure}
  \centering
  \includegraphics[width=0.5\textwidth]{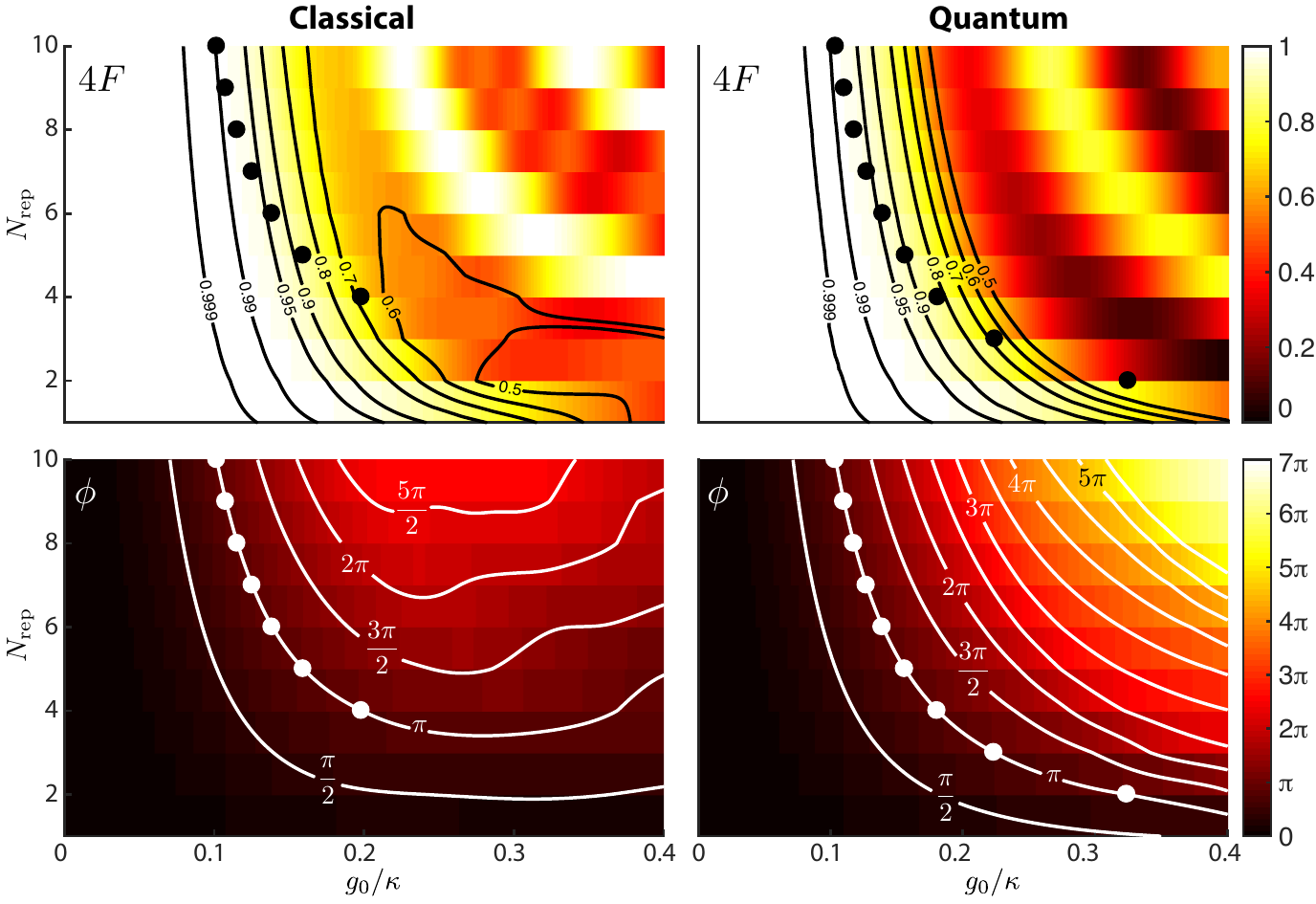}
  \caption{Comparison between semiclassical model and full quantum simulation. The black contours outline different gate fidelities, while the white contours show the generated phase shift. The black and white dots are the $pi$-points in parameter space showing the values of $g_0/\kappa$ and $\Nbnc$ required for a $\pi$ phase gate. The primary source for loss of fidelity and phase shift are captured by our semiclassical model, while other  contributions are revealed in the full quantum simulation. It is worth noting that the agreement between the simple classical calculation and the quantum model becomes worse at larger coupling $g_0/\kappa$ and that quantum model seems to lead to larger phase shifts.}
  \label{fig:si_classical_quantum}
\end{figure}

We can estimate this effect in a semiclassical model. A photon bouncing off a mechanical system that is in a coherent state $\ket{\beta}$ (instead of in its ground state) can be described by a displacement operator in mechanical phase space:
\bea
U_b (\beta) = \exp \left(\frac{-4ig_0/\kappa}{1 + 4(g_0/\kappa)^2(\beta + \beta^*)^2} (b+b^\dagger) \right)
\eea
The reason is that the cavity shift induced by the mechanical displacement ($\beta+\beta^\ast$) reduces the momentum kick induced by the interaction with the photon.
After the interaction with first photon, the mechanical changes from being in its ground state $\ket{0}$ to $U_b(0)\ket{0} = \ket{-ir}$ which is also a coherent state, where $r = 4g_0/\kappa$. After $T_m/4$ free evolution, the mechanical state becomes $\ket{-r}$ -- the momentum kick induced by the photon becomes a displacement. Then the interaction with the second photon  changes the mechanical state to $U_b(-r)\ket{-r} = e^{i\phi_1}\ket{-r \left(1 + \frac{i}{1 + r^4}\right)}$, where $\phi_1 = \frac{r^2}{1 + r^4}$. Continuing this calculation until the end of one repetition of the protocol leads to a mechanical final state $\ket{\beta_r}$ with
\bea
\beta_r =  \frac{r^5}{(1+r^4)^2 + r^4} + ir \left( \frac{1}{1+r^4} - \frac{1}{1 + r^2\left( \frac{r^5}{(1+r^4)^2 + r^4} \right)^2}  \right). \nonumber
\eea
When $r<1$, $\beta_r \approx r^5 (1+i)$. To compare the contribution of this effect to the fidelity of our gate, we numerically keep track of the evolution of the mechanical state for $\ket{11}$ input state and calculate the fidelity as well as the phase shift for each repetition of the protocol. As shown in Figure~\ref{fig:si_classical_quantum}, the semiclassical effect is a major part of the contribution to the full quantum results. Here the fidelity is defined as $|\langle 0 | \beta_r \rangle|^2 = e^{-|\beta_r|^2}$.

\textbf{Fidelity change caused by another mechanical mode.} The Hamiltonian of system with other mechanical modes will have other terms $ \xi (t) a^\dagger a $ where $\xi (t) = g_p x_p(t)$ is the frequency jitter due to a parasitic mechanical mode. We assume there is only one other mechanical modes, that during a photon interaction, $\xi (t)$ is approximately constant, and model this parasitic mode classically. Then the extra phase shift can be estimated as $ \exp (-i \xi \int_{\tau} a^\dagger (t) a(t) \text{d}t) = \exp (-i4\xi/\kappa) $. With an entangled photonic state input $ \ket{+} = \frac{1}{\sqrt{2}}(\ket{10} + \ket{01}) $, after a perfect phase gate, since $\ket{10}$ and $\ket{01}$ should have the same phase, the state will remain $\ket{+}$. However, since the parasitic mechanical mode incurs a random phase shift, the final state becomes $\ket{10} + e^{i\theta}\ket{01}$ with $\theta = \frac{4}{\kappa} (\xi_1 + \xi_3 - \xi_2 - \xi_4)$. Here $\xi_1 (\xi_2)$ and $\xi_3 (\xi_4)$ are related to the random phase shift obtained by the $\ket{10} (\ket{01})$ during its two bounces on the mechanical oscillator. We model $\xi_i$ as independent random variables obeying normal distribution $\mathcal{N}(0, \sigma^2)$. Therefore the distribution for $\theta$ is $\theta \sim \mathcal{N}(0, 4(\frac{4\sigma}{\kappa})^2)$. For a given $\theta$ the fidelity is $F(\theta) = \frac{1 + \cos \theta}{2}$, and on average the fidelity is
\bea
F =& \int \text{d}\theta f(\theta) F(\theta) = \int \text{d}\theta f(\theta) \frac{1 + \cos \theta}{2} \\
=& \frac{1}{2}(1 + e^{- 32\sigma^2 /\kappa^2}) \approx 1 - 16\sigma^2 / \kappa^2
\eea
where $f(\theta)$ is the probability distribution of $\theta$. Since the phase fluctuation can be estimated by $ \sigma = \frac{g_p}{\omega_p}\sqrt{\frac{k_B T}{m_p}} $, we arrive at an expression bounding $g_p$ and the temperature $T$ for a given required fidelity.

%

\clearpage

\end{document}